\RequirePackage[switch]{lineno_babar_revtex}                                                                                                                                           
\documentclass[twocolumn,showpacs,aps,prl]{revtex4}   

\catcode`\@=11\relax
\def\@makecol{%
 \setbox\@outputbox\vbox{%
  \boxmaxdepth\@maxdepth
 \protected@write\@auxout{}{% 
 \string\@LN@col{\@ifnum{\pagegrid@cur=\@ne}{1}{2}}
      }%
  \@tempdima\dp\@cclv
  \unvbox\@cclv
  \vskip-\@tempdima
 }%
 \xdef\@freelist{\@freelist\@midlist}\global\let\@midlist\@empty
 \@combinefloats
 \@combineinserts\@outputbox\footins
  \set@adj@colht\dimen@
  \count@\vbadness
  \vbadness\@M
  \setbox\@outputbox\vbox to\dimen@{%
   \@texttop
   \dimen@\dp\@outputbox
   \unvbox\@outputbox
   \vskip-\dimen@
   \@textbottom
  }%
  \vbadness\count@
 \global\maxdepth\@maxdepth
}%
\def\balance@two#1#2{%
\outputdebug@sw{{\tracingall\scrollmode\showbox#1\showbox#2}}{}%
 \setbox\@ne\vbox{%
  \@ifvoid#1{}{%
   \unvcopy#1\recover@footins
   \@ifvoid#2{}{\marry@baselines}%
  }%
  \@ifvoid#2{}{%
   \unvcopy#2\recover@footins
  }%
 }%
 \dimen@\ht\@ne\divide\dimen@\tw@
 \dimen@i\dimen@
 \vbadness\@M
 \vfuzz\maxdimen
 \loopwhile{%
  \dimen@i=.5\dimen@i
  \outputdebug@sw{\saythe\dimen@\saythe\dimen@i\saythe\dimen@ii}{}%
  \setbox\z@\copy\@ne\setbox\tw@\vsplit\z@ to\dimen@
  \setbox\z@ \vbox{%
 \protected@write\@auxout{}{% 
 \string\@LN@col{\@ifnum{\pagegrid@cur=\@ne}{1}{2}}
      }%
   \unvcopy\z@
   \setbox\z@\vbox{\unvbox\z@ \setbox\z@\lastbox\aftergroup\vskip\aftergroup-\expandafter}\the\dp\z@\relax
  }%
  \setbox\tw@\vbox{%
   \unvcopy\tw@
   \setbox\z@\vbox{\unvbox\tw@\setbox\z@\lastbox\aftergroup\vskip\aftergroup-\expandafter}\the\dp\z@\relax
  }%
  \dimen@ii\ht\tw@\advance\dimen@ii-\ht\z@
  \@ifdim{\dimen@i>.5\p@}{%
   \advance\dimen@\@ifdim{\dimen@ii<\z@}{}{-}\dimen@i
   \true@sw
  }{%
   \@ifdim{\dimen@ii<\z@}{%
    \advance\dimen@\tw@\dimen@i
    \true@sw
   }{%
    \false@sw
   }%
  }%
 }%
 \outputdebug@sw{\saythe\dimen@\saythe\dimen@i\saythe\dimen@ii}{}%
\@ifdim{\ht\z@=\z@}{%
\@ifdim{\ht\tw@=\z@}{%
\true@sw
}{%
\false@sw
}%
}{%
\true@sw
}%
{%
}{%
\ltxgrid@info{Unsatifactorily balanced columns: giving up}%
\setbox\tw@\box#1%
\setbox\z@ \box#2%
}%
 \setbox\tw@\vbox{\unvbox\tw@\vskip\z@skip}%
 \setbox\z@ \vbox{\unvbox\z@ \vskip\z@skip}%
 \set@colroom
\dimen@\ht\z@\@ifdim{\dimen@<\ht\tw@}{\dimen@\ht\tw@}{}%
\@ifdim{\dimen@>\@colroom}{\dimen@\@colroom}{}%
 \outputdebug@sw{\saythe{\ht\z@}\saythe{\ht\tw@}\saythe\@colroom\saythe\dimen@}{}%
\setbox#1\vbox to\dimen@{\unvbox\tw@\unskip\raggedcolumn@skip}%
\setbox#2\vbox to\dimen@{\unvbox\z@ \unskip\raggedcolumn@skip}%
\outputdebug@sw{{\tracingall\scrollmode\showbox#1\showbox#2}}{}%
}%
\catcode`\@=12\relax

\usepackage{graphicx}
\usepackage{verbatim}
\usepackage{dcolumn}
\usepackage{amsmath}
\usepackage{epsfig}
\usepackage{subfigure}

\newcommand{\BaBarYear}      {18}
\newcommand{\BaBarNumber}    {010}
\newcommand{\BaBarType}      {PUB}  
\newcommand{\SLACPubNumber}  {17455}

\input babarsym.tex

\def\epem{e^+e^-}
\def\knunu      {\ensuremath{\B\to K^{(*)}\nu\nub}\xspace}
\def\Blpnn      {\ensuremath{\Bm\to \Lambda \antiproton \nu\nub}\xspace}
\def\Lambdatoppi {\ensuremath{\Lambda \to \proton \pim}\xspace}
\def\Btag{\ensuremath{B_{\rm tag}}\xspace}  
\def \Bsig{\ensuremath{B_{\rm sig}}\xspace}
\def \lumi{\ensuremath{424\invfb}\xspace} 
\def \NBB{\ensuremath{(471 \pm 3) \times 10^{6}}\xspace} 
\def \BFresult{\ensuremath{\BR (\Blpnn) =  (0.4 \pm 1.1\textrm{ (stat.)} \pm 0.6\textrm{ (sys.)})\times 10^{-5}}\xspace} 
\def \BFlimit{\ensuremath{\BR (\Blpnn) < 3.0\times 10^{-5}}\xspace} 
\def \BFsignal{\ensuremath{\BR (\Blpnn) =  0.4 \times 10^{-5}}\xspace}

\def\epsilonLimLambda {\ensuremath{7.4}\xspace}

\begin{document}
\pagestyle{plain}

\begin{flushleft}
\babar-\BaBarType-\BaBarYear/\BaBarNumber \\
SLAC-PUB-\SLACPubNumber\\
\end{flushleft}

\title{\vskip 20pt{\large \boldmath Search for $\Blpnn$ with the \babar{} experiment}}

\author{J.~P.~Lees}
\author{V.~Poireau}
\author{V.~Tisserand}
\affiliation{Laboratoire d'Annecy-le-Vieux de Physique des Particules (LAPP), Universit\'e de Savoie, CNRS/IN2P3,  F-74941 Annecy-Le-Vieux, France}
\author{E.~Grauges}
\affiliation{Universitat de Barcelona, Facultat de Fisica, Departament ECM, E-08028 Barcelona, Spain }
\author{A.~Palano}
\affiliation{INFN Sezione di Bari and Dipartimento di Fisica, Universit\`a di Bari, I-70126 Bari, Italy }
\author{G.~Eigen}
%\author{B.~Stugu}
\affiliation{University of Bergen, Institute of Physics, N-5007 Bergen, Norway }
\author{D.~N.~Brown}
\author{Yu.~G.~Kolomensky}
%\author{M.~J.~Lee}
%\author{G.~Lynch}
\affiliation{Lawrence Berkeley National Laboratory and University of California, Berkeley, California 94720, USA }
\author{M.~Fritsch}
\author{H.~Koch}
\author{T.~Schroeder}
\affiliation{Ruhr Universit\"at Bochum, Institut f\"ur Experimentalphysik 1, D-44780 Bochum, Germany }
% ...
\collaboration{The \babar\ Collaboration}
\noaffiliation


\begin{abstract}

A search for the rare flavor-changing neutral current process $\Blpnn$ using data from the \babar{} experiment has been performed. A total of $\lumi$ of $e^+e^-$ collision data collected at the center-of-mass energy of the \FourS resonance is used in this study, corresponding to a sample of $\NBB$ $\BB$ pairs. Signal $\Blpnn$ candidates are identified by first fully reconstructing a $\Bp$ decay in one of many possible exclusive decays to hadronic final states, then examining detector activity that is not associated with this reconstructed $\Bp$ decay for evidence of a signal $\Blpnn$ decay. The data yield is found to be consistent with the expected background contribution under a null signal hypothesis, resulting in an upper limit of $\BFlimit$ at the $90\%$ confidence level.
\end{abstract}

\maketitle

\setcounter{footnote}{0}

\begin{figure}[ht]
\begin{center}
\includegraphics[width=6cm]{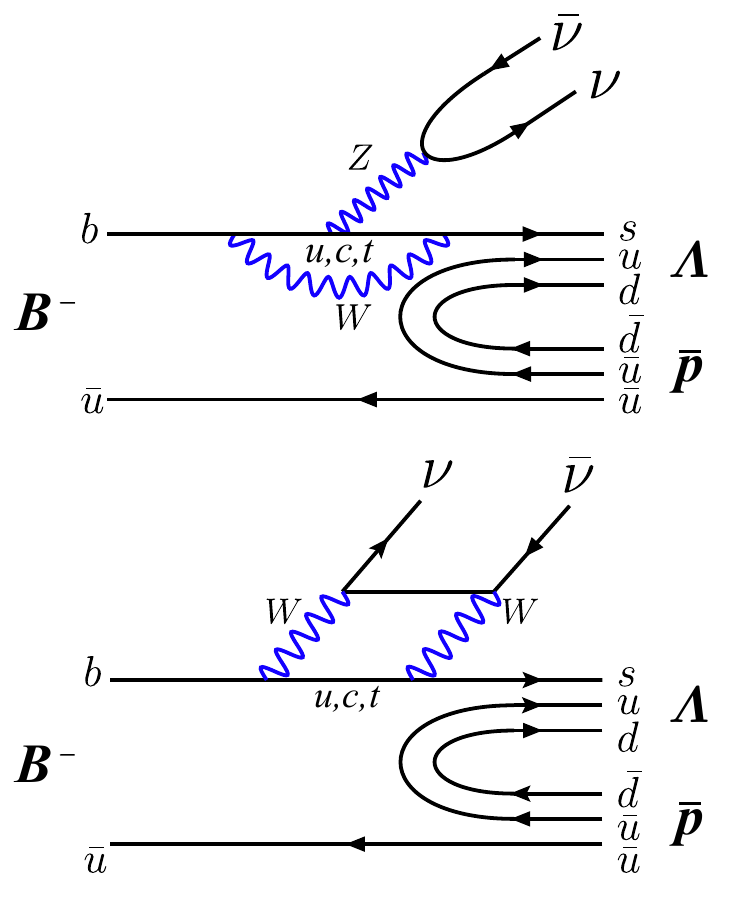}
\caption{Lowest order diagrams of $\Blpnn$ in the SM. Adapted from Ref.~\cite{geng}.}
\label{fig1}
\end{center}
\end{figure}

Flavor-changing neutral current (FCNC) processes are suppressed in the standard model (SM) of particle interactions, first appearing at one-loop level.  Consequently, new physics contributions could result in potentially measurable deviations from SM predictions. The process $\Blpnn$ ({\it CP} conjugate processes are implied throughout this paper) is the baryonic analog of $\knunu$, occurring in the SM via a FCNC $b \to s \nu \nub$ transition through  $Z$-penguin or $W$-box processes (see Fig.~\ref{fig1}).   The branching fraction is predicted to be $\BR(\Blpnn) = (7.9 \pm 1.9 ) \times 10^{-7} $~\cite{geng}.  Although $\knunu$ has previously been studied at $B$ factory experiments \cite{knunuPaper, Belle_Knunu},  it is challenging due to the presence of two (unobserved) neutrinos in the final state and current measurements leave room for new physics~\cite{straub}. By comparison, the presence of two baryons in the final state of $\Blpnn$ provides stronger background rejection. This paper presents the first search for the decay $\Blpnn$, using data recorded by the $\babar$ experiment at the PEP-II energy-asymmetric $e^+e^-$  collider.  These data were collected at the $\FourS$ resonance, representing an integrated luminosity of $\lumi$~\cite{lumi_paper}, corresponding to $\NBB$ $\BB$ pairs~\cite{Bcount}. 

The $\babar$ detector is described in detail in Refs.~\cite{Babar1, Babar2}. The charged-particle tracking system comprises a five-layer silicon vertex tracker and a 40-layer cylindrical drift chamber. A $1.5$~T magnetic field produced by a superconducting solenoid enables momentum measurement of charged particles. Identification of (anti)protons and other charged particles is based on measurement of the specific ionization, ${\rm d}E/{\rm d}x$, in the tracking detectors, combined with information from the electromagnetic calorimeter and Cherenkov-photon angle information from an array of fused silica quartz bars. Energy and position measurements for photons are provided by an electromagnetic calorimeter comprising 6580 CsI(Tl) crystals arrayed as a cylindrical central barrel and a conical forward endcap.  

Simulated Monte Carlo (MC) event samples are used to develop the signal selection and to estimate the selection efficiency. Studies of background channels are based on samples of  simulated events representing $\BB$ production at $\FourS$, and continuum production of $\epem\to\qqbar$ and $\epem\to\tautau$. The $\qqbar$ simulation is separated into $\ccbar$ and light quark ($\uubar, \ddbar, \ssbar$) samples. The $\BB$ samples are produced using EvtGen~\cite{evtgen}, while JETSET~\cite{jetset} is used for generation and hadronization of continuum backgrounds, with EvtGen handling decays.   KK~\cite{KKMC} is used for \tautau generation,  with Tauola~\cite{tauola} handling $\tau$ decays.
The detector simulation uses GEANT4~\cite{Geant4}.  The $\BpBm$, $\BzBzb$, and $\ccbar$ simulation samples correspond to an integrated luminosity ten times that of data, and the other samples are four times that of data.  A dedicated  $\Blpnn$,  $\Lambdatoppi$  signal MC sample of $4.053 \times 10^6$ events is used for efficiency and optimization studies.    These events are generated according to a phase-space model, but are adapted to the form factor model described in Ref.~\cite{geng} by applying a re-weighting of the di-baryon invariant mass, $m_{\Lambda \antiproton}$, at the analysis level.

Because the decay $\Blpnn$ has two undetected neutrinos, it cannot be fully reconstructed from its final state particles.  Instead, by reconstructing the hadronic decay of one of the $\B$ mesons in $\FourS \to \BB$ events, referred to as the ``tag $\B$'' ($\Btag$), all remaining particles in the event can then be inferred to be daughters of the other $\B$, referred to as the ``signal $B$'' ($\Bsig$). The 4-vector of the $\Bsig$ can be calculated from the $\Btag$ momentum vector, $\vec p^{\mkern 3.5mu*}_{\Btag}$, and the known CM energy, $E^{*}_ {\rm CM}$: $|\vec p^{\mkern 3.5mu*}_{\Bsig}| = \sqrt{(E^{*}_{ \rm CM}/2)^2 - m^2_{\B}}$, where $\vec p^{\mkern 3.5mu*}_{\Bsig}$ is the three-momentum vector of the \Bsig, $E^{*}_{\rm CM}$ is the CM energy, and $m_{B}$ is the \B meson mass, with the direction of $\vec p^{\mkern 3.5mu*}_{\Bsig}$ defined to be opposite that of  $\vec p^{\mkern 3.5mu*}_{\Btag}$,  where asterisks indicate quantities in the CM frame.   The missing momentum four-vector, $p^{*}_{\rm miss}$, is determined by subtracting the CM four-momentum of all identified particles that are not used in the reconstruction of the $\Btag$ from that of $\Bsig$.   Since the $\Btag$ has been fully reconstructed, all missing momentum in the event is attributable to the $\Bsig$ candidate. This method has been used in previous $\babar$ analyses, e.g.\ Refs.~\cite{Ktautau, knunuPaper, lnugamma}.
 
\begin{figure}
\begin{center}
\includegraphics[width=\columnwidth]{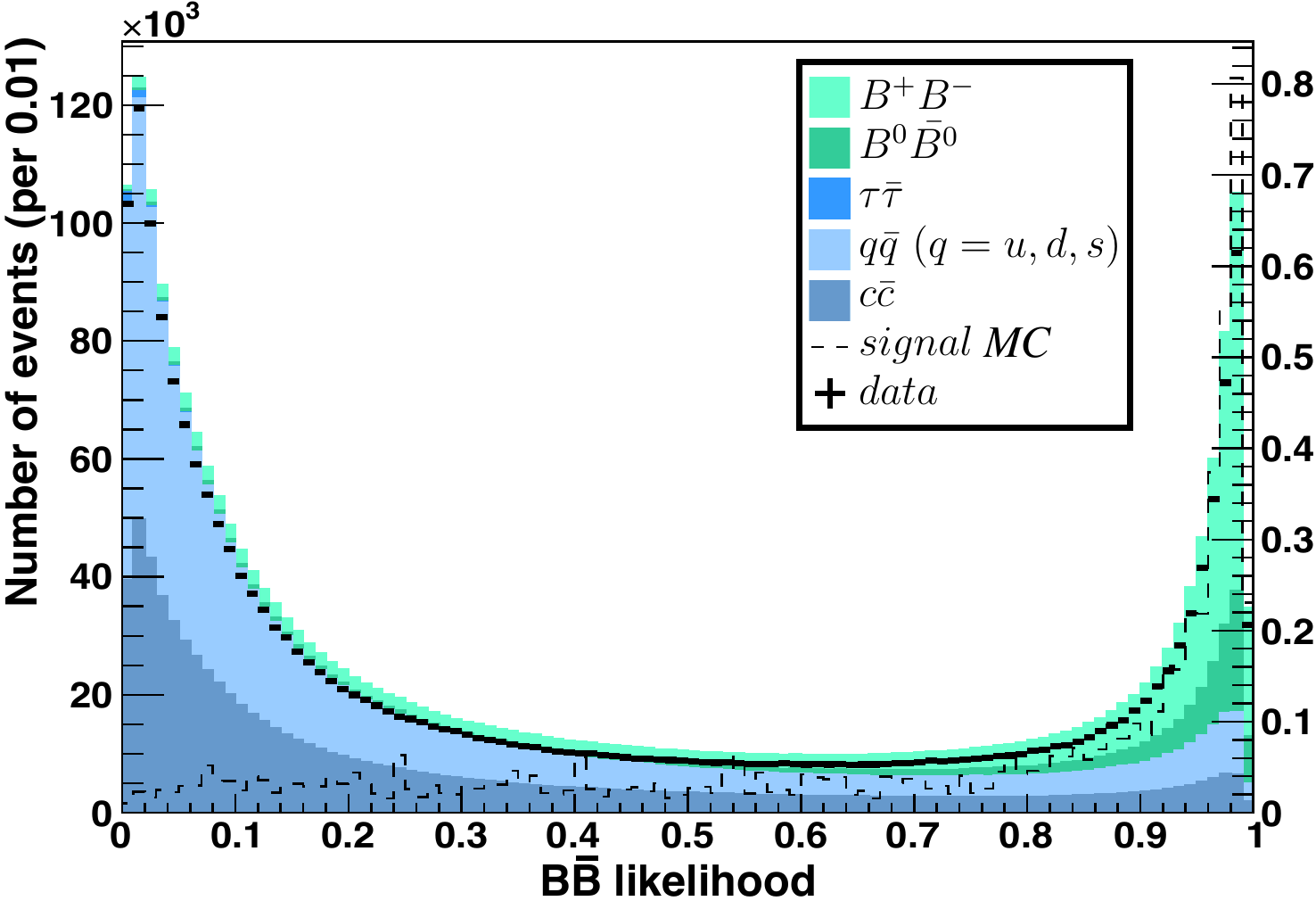}
\caption{Output of the \BB likelihood selector,  $\mathcal{ L}_{\BB}$, for data (points with error bars) and background MC (stacked, shaded histograms) normalized to the data luminosity, for events with a reconstructed $\Btag$ with $5.27~\gevcc<\mes<5.29~\gevcc$. The expected distribution for simulated $\Blpnn$ events is also shown overlaid for a branching fraction of $0.4 \times 10^{-5}$ (dashed line), with yields per 0.01 given by the $y$-axis on the right-hand side. }
\label{fig2}
\end{center}
\end{figure}

\begin{figure}
\begin{center}
\includegraphics[width=\columnwidth]{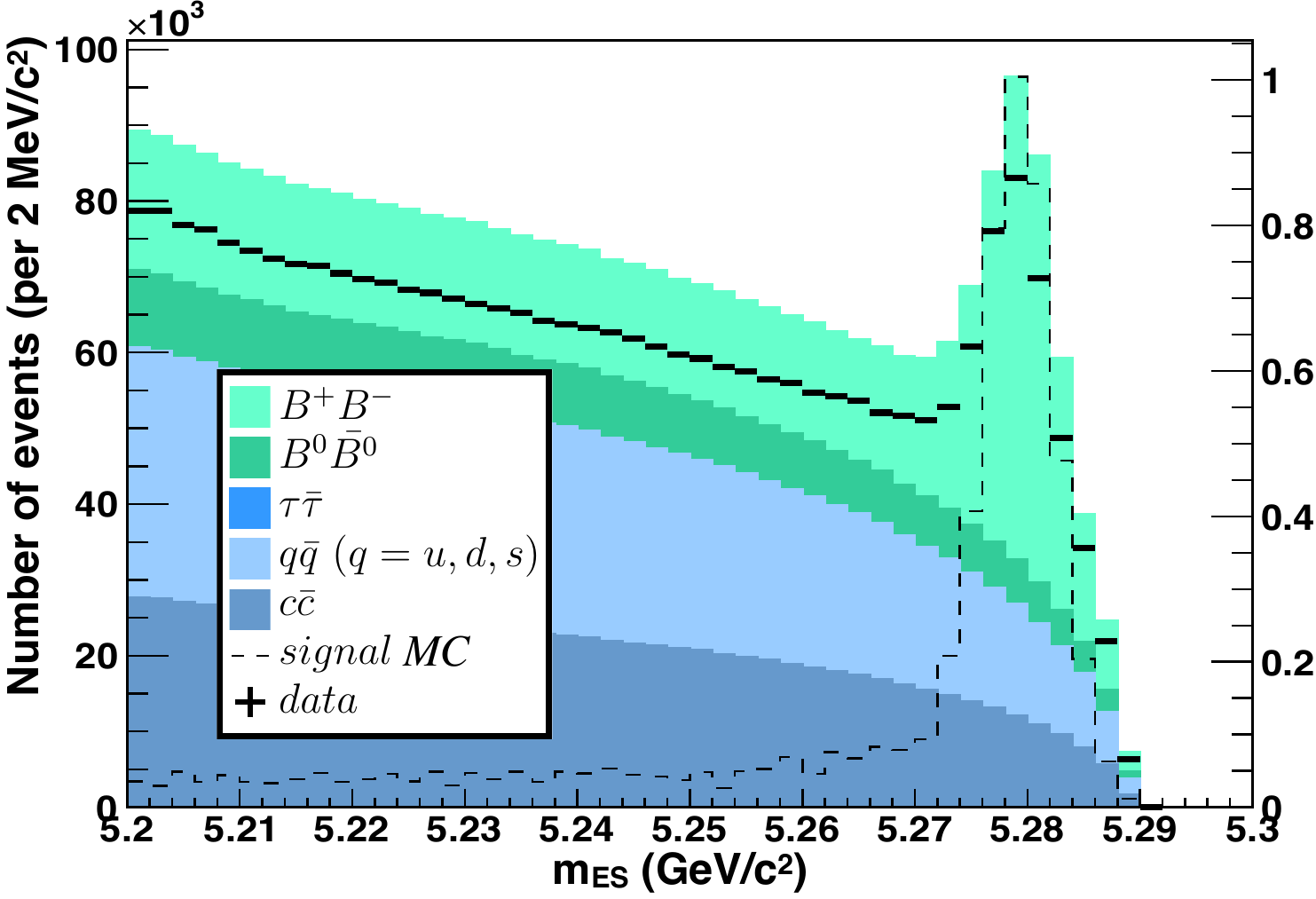}
\caption{The $\mes$ distribution for data (points with error bars) and background MC (stacked, shaded histograms) normalized to the data luminosity, for events which satisfy the continuum suppression criterion $\mathcal{ L}_{\BB} > 0.35$.  The expected distribution for simulated $\Blpnn$ events is also shown overlaid for a branching fraction of $0.4 \times 10^{-5}$ (dashed line), with event yields per 2 \mevcc given by the $y$-axis on the right-hand side. }
\label{fig3}
\end{center}
\end{figure}

The reconstruction of $\Btag$ candidates considers \B decays into a large number of possible hadronic decay modes,  \B\to$S X$, 
where $S$ is a ``seed'' meson, and $X$ is an hadronic system comprising up to five kaons or pions with total charge 0 or $\pm 1$.  Both neutral and charged $\Btag$ candidates are reconstructed, but only $\Bpm$ candidates are retained for this study. The seed meson can be $D^{(*)0}$, $D^{(*)\pm}$,  $D^{*\pm}_{s}$, or $\jpsi$. The $D$ meson seeds are reconstructed as: $D^+\to\KS\pip$, $\KS\pip\piz$, $\KS\pip\pim\pip$, $ K^-\pip\pip$, $K^-\pip\pip\piz$, $K^{+}\Km\pip$, and $K^{+}\Km\pip\piz$; $\Dz\to K^-\pip$, $K^-\pip\piz$, $K^-\pip\pim\pip$, $\KS\pip\pim$, $\KS\pip\pim\piz$, $K^{+}\Km$, $\pip\pim$, $\pip\pim\piz$, and $\KS\piz$; $D^{*+}\to \Dz\pip$, and $\Dp\piz$; $\Dstarz\to \Dz\piz$, and $\Dz\g$. The $ \Dss$ seed decay consists of $D_s^{*+}\to D_s^+\g$; $D_s^+\to\phi\pip$, and $\KS K^{+}$.  The $J/\psi$ seed is reconstructed via $\epem$ and $\mu^+\mu^-$.  $\pi^0 \to \gamma \gamma$, $\KS \to  \pip\pim$, and $\phi \to  K^{+}K^{-}$ are reconstructed.
A kinematic fit is applied, which imposes vertex and particle mass constraints on the candidates. The resulting seed candidates are then combined with kaons or pions to create $\Btag$ candidates. 
Two kinematic variables are used to define these candidates: $\mes =\sqrt{(s/2 + \vec{p}_{B_{\rm tag}}\cdot \vec{p}_0)^2/E^2_0 - \vec{p}^{\,2}_{B_{\rm tag}}},$ and $\Delta E=E^{*}_{\rm CM}/2-E^{*}_{ \Btag},$ where $E_0$ and $\vec{p}_0$ are the energy and momentum of the $e^+e^-$ system in the lab frame, and $\sqrt{s}$ is the energy of the $e^+e^-$ system in the CM frame. The $\Btag$ candidates are selected by requiring $-0.12~\gev<\Delta E<0.12~\gev$ and $5.20~\gevcc<\mes<5.30~\gevcc$. If multiple candidates are present in an event, they are ranked based on the value of the reconstructed seed candidate mass with respect to the nominal mass of this particle, and the magnitude of $\Delta E$. Only a single $\Btag$ candidate per event is retained.   Individual $\Btag$  modes with a measured high level of combinatorial background are subsequently excluded.  The overall tagging efficiency is sub-percent \cite{BFactories}. 
Correctly-reconstructed $\Btag$ candidates contribute to a peak in the $\mes$ distribution near the $\B$ meson mass.  The interval $5.27~\gevcc<\mes<5.29~\gevcc$ is defined as the signal region, and the interval $5.20~\gevcc<\mes<5.26~\gevcc$ as the sideband region. Continuum processes, from non-resonant $e^+e^- \to q \bar{q}$, and incorrectly reconstructed $\BB$ decays result in a substantial combinatorial background in both the signal and sideband regions.  The continuum background is suppressed using a multivariate likelihood comprising six inputs which distinguish between comparatively jet-like non-resonant processes and more isotropic decay topologies of $\FourS \to \BB$.
The inputs are: the ratio of the second and zeroth Fox-Wolfram moments~\cite{wolfram}, calculated using all reconstructed charged tracks and calorimeter clusters in the event;
the event thrust vector, the sum of the magnitudes of the momenta of all tracks and clusters projected onto the thrust axis, where the thrust axis is the axis that maximises the projection, and where the thrust vector is normalised with respect to the sum of the magnitudes of the momenta; the magnitude of the projection of the thrust vector onto the $z$-axis; the cosine of the angle between the $\Btag$ direction and the $z$-axis; the cosine of the angle between the event's missing momentum vector and the $z$-axis; the cosine of the angle between the thrust axes of the decay daughters of the $\Btag$ and of the $\Bsig$. These quantities are computed in the CM frame. The selector output, $\mathcal{ L}_{\BB}$, is shown in Fig.~\ref{fig2}.   Events with  $\mathcal{ L}_{\BB} > 0.35$ are retained. This requirement rejects $76\%$ of continuum background events and $16\%$ of \BB background events while retaining $82\%$ of signal events. The \mes distribution of events selected by this criterion is shown in Fig.~\ref{fig3}.

\begin{figure}
\begin{center}
\includegraphics[width=\columnwidth]{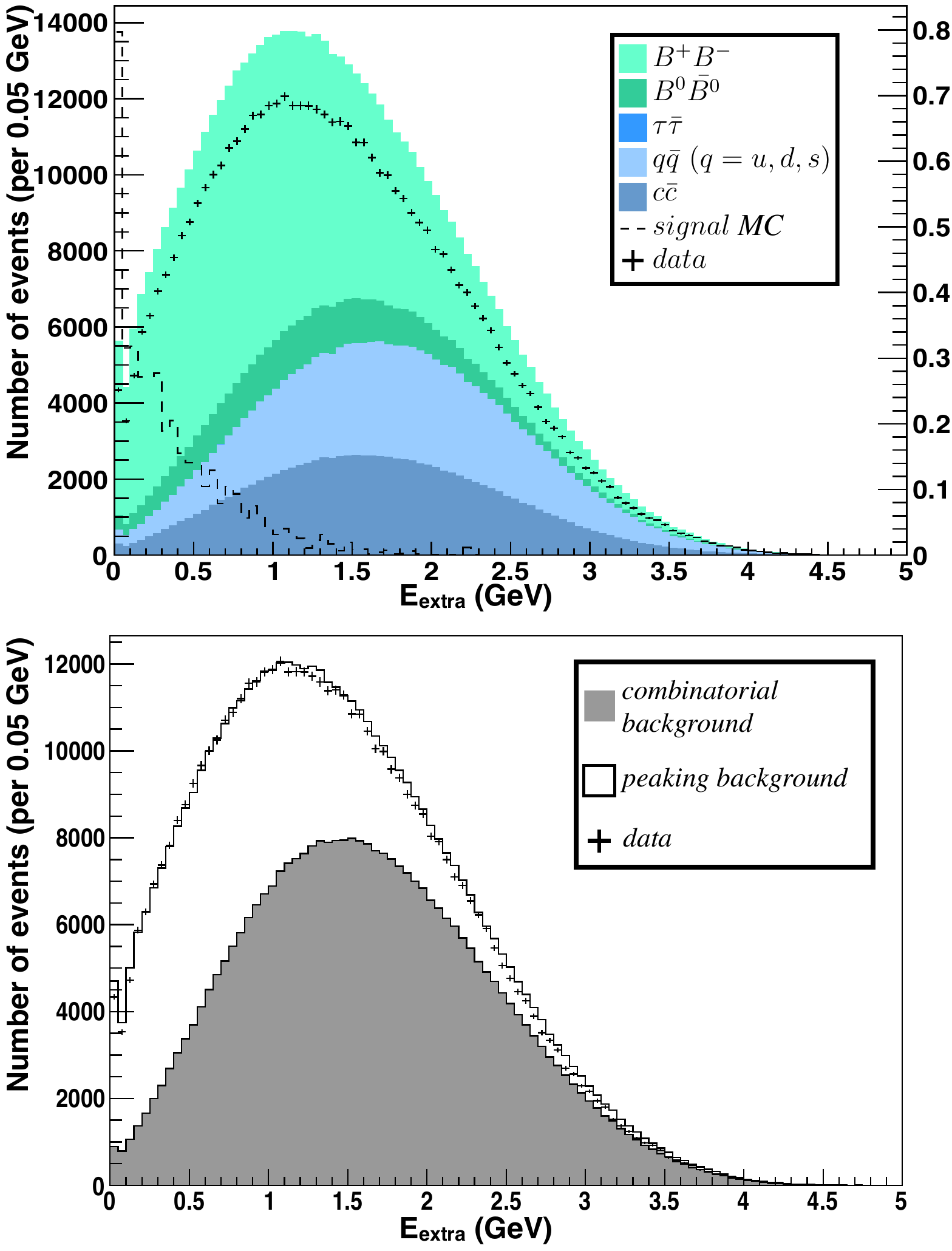}
\caption{Distribution of $E_{\rm extra}$, calculated in the CM frame, in data and MC before (top) and after (bottom) application of the MC correction procedure for events with a reconstructed $\Btag$ with $\mes$ within the signal region. In the upper plot, data are shown as points with error bars, background MC is shown as stacked, shaded histograms. The expected distribution for simulated $\Blpnn$ events is shown overlaid for a branching fraction of $0.4 \times 10^{-5}$ (dashed line), with yields given by the $y$-axis on the right-hand side.  In the lower plot the shaded region is the sideband data scaled by $R_{\rm side}$, the unshaded region is the $\mes$ peaking component of the $\BpBm$ MC scaled by $C_{\rm peak}$.}
\label{fig4}
\end{center}
\end{figure}

The $\Blpnn$ candidates are identified by considering all activity in the detector which is not associated with the reconstructed $\Btag$.   Since only the $\Lambdatoppi$ decay mode is considered in this analysis, $\Bsig$  candidates are required to possess exactly three charged tracks, with total charge opposite that of the $\Btag$. Signal events typically contain several low-energy clusters in the calorimeter from hadronic shower fragments, bremsstrahlung, or beam-related sources. Physics backgrounds, however, frequently also produce higher energy clusters from $\pi^0$  decays and similar processes.  These backgrounds are suppressed by requiring $E_{\rm extra} <400$~MeV, where $E_{\rm extra}$ is the total CM-frame energy of $\Bsig$ clusters which have lab-frame energy exceeding $50$~MeV; see Fig.~\ref{fig4} (top).

\begin{figure}
\begin{center}
\includegraphics[width=\columnwidth]{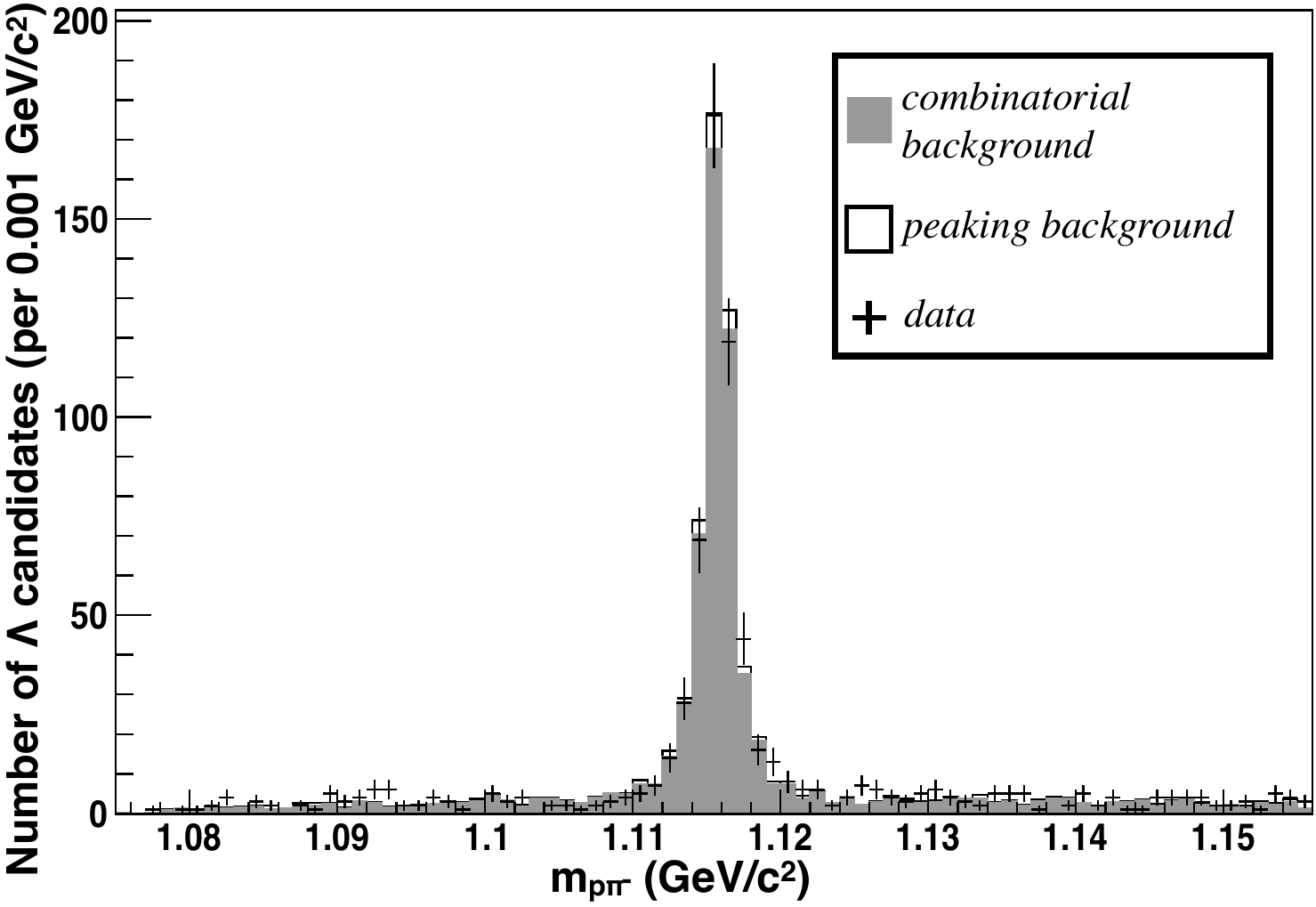}
\caption{The $\proton \pi^-$ invariant mass in events with a reconstructed $\Btag$ with $\mes$ within the signal region, with three charged tracks satisfying the (anti)proton selection and DOCA requirements. Data are shown as points with error bars, the shaded region is the sideband data scaled by $R_{\rm side}$, the unshaded region is the $\mes$ peaking component of the $\BpBm$ MC scaled by $C_{\rm peak}$.}
\label{fig5}
\end{center}
\end{figure}

The background MC does not accurately reproduce the event yield in data at this point in the selection. This deficiency has been observed in previous $\babar$ analyses \cite{Ktautau, knunuPaper, lnugamma} and is understood to be due to a combination of inaccurate branching fractions and modelling of $\Btag$ reconstruction efficiencies in the simulation. A two step procedure is applied to correct this. Events in the $\mes$ signal region can be divided into correctly reconstructed (``peaking") and combinatorial (``non-peaking") components. The non-peaking component in the signal region is determined from data by extrapolation of the $\mes$ sideband data into the signal region. The shape of this distribution is obtained from background MC, and is characterized by the quantity $R_{\rm side}$, the ratio of the MC non-peaking background yield in the signal region to the yield in the sideband-region. After the signal selection described above, $R_{\rm side}$ is evaluated as $0.215 \pm 0.001$, where the uncertainty is due to MC statistics. Scaled sideband data are then substituted for combinatorial MC in the $\mes$ signal region when studying distributions of selection variables. Once the combinatorial contribution in the signal region has been determined, it is combined with the subset of $\BpBm$ MC in which a $\Btag$ has been correctly reconstructed, resulting in the peaking contribution in the $\mes$ distribution. This peaking MC contribution is scaled by a factor $C_{\rm peak} = 0.819 \pm 0.006$ to match data. Following this procedure, excellent agreement is observed in all kinematic variables used in this analysis, e.g. Figs.~\ref{fig4} -~\ref{fig5}. 
As the quantity  $C_{\rm peak}$ represents a global correction to the $\Btag$ yield, it is also applied to the signal efficiency. The reconstruction efficiency for $\FourS$ events containing a $\Blpnn$ decay is estimated to be approximately $0.07\%$, after requiring that events possess a $\Btag$ with $\mes$ in the signal region and satisfy the signal selection described above. The remainder of the event selection optimization is performed ``blind'', i.e., without knowledge of the data yield in the signal region until the selection procedure has been finalized.

Decays of $\Bsig$ candidates are expected to contain a proton-antiproton pair and a single charged pion, where the (anti)proton with the same charge as the $\Btag$ is presumed to be the daughter of the $\Lambda$.    Tight (anti)proton particle identification criteria are applied to the baryon candidate tracks; no pion identification requirement is imposed on the third track. The (anti)proton selectors have an efficiency of approximately 95\% within the momentum range relevant to this analysis~\cite{Babar2}. A kinematic fit is imposed on the $\Lambda$ daughter tracks, applying pion and proton mass hypotheses and fitting the $\Lambda$ vertex, including a constraint that the $\Lambda$ originates within a $\B$ meson flight length of the event vertex. The three tracks are required to have a DOCA ordering consistent with a $\Blpnn$ signal event, where DOCA is defined as the extrapolated distance of closest approach of a reconstructed track to the nominal event vertex.  The $\antiproton$ that is the daughter of the $\Bsig$ originates from near the interaction point and so usually has the smallest DOCA.
The two $\Lambda \to \proton \pi^-$ decay daughters typically do not point to the interaction point, with the $\proton$ that is the daughter of the $\Lambda$ usually having a smaller DOCA than the $\pi^-$.   The resulting $\proton \pi^-$ invariant mass distribution, without any $\mathcal{ L}_{\BB}$ or $E_{\rm extra}$ requirements, is shown in Fig~\ref{fig5}.  The $\Lambda$ candidates are selected by requiring  $1.112~\gevcc < m_{\proton\pi^-} < 1.120~\gevcc$. Following this selection, background events are almost entirely real $\Lambda$ baryons from $\qqbar$ continuum sources.

A simultaneous optimization of the  $\mathcal{ L}_{\BB}$ and  $E_{\rm extra}$ selection criteria is performed, with the expected branching fraction limit in the absence of signal used as the figure of merit. This optimization yields the selection criteria values presented previously. The signal efficiency is estimated to be $(0.034 \pm 0.001~{\rm (stat.)})\%$.   The background yield is determined by combining the peaking background from $\BpBm$ MC with the combinatorial background estimated from the $\mes$ sideband, yielding $2.3 \pm 0.7~{\rm (stat.)}$ events. The dominant contribution of $1.7 \pm 0.6~{\rm (stat.)}$ arises from combinatorial background sources.

Systematic uncertainties arise in the determination of the signal efficiency and background yield. The combinatorial background yield is determined from data by extrapolation of the sideband into the $\mes$ signal region. However, the shape of the combinatorial background distribution impacts the peaking yield correction and hence $C_{\rm peak}$ is anti-correlated with $R_{\rm side}$. Consequently, the relevant systematic uncertainty is due to the extrapolation of the yield of combinatoric events in the $\mes$ sideband to the signal region. The ratio $R_{\rm side}$ is obtained from non-peaking background MC ($\qqbar$, $\ccbar$, \tautau, $\BzBzb$, and non-peaking $\BpBm$) and its value depends on the relative mix of the continuum and $\BB$ due to the difference in shape in the predicted $\mes$ distributions of these two components. An uncertainty of $17\%$ on background yield and $16\%$ on signal efficiency is obtained by varying the shape of the \mes distribution between that given by $\BB$ and continuum MC, and determining the impact on the resulting signal efficiency and background estimates.

The signal MC is produced using a phase-space model, which is subsequently weighted into the model of Ref.~\cite{geng}, based on the $m_{\Lambda \antiproton}$ distribution. The impact of this weighting on the signal efficiency is evaluated by modifying the weighting scheme to include the other kinematic quantities $m_{\nunub}$ and $\theta_{B, L}$ defined in that paper. A systematic uncertainty of $9.6\%$ is assigned.

MC modelling of variables used in the signal selection impact both the signal efficiency and the background determination. The impact of (anti)proton particle identification is evaluated using standard $\babar$ procedures~\cite{Babar2} for the relevant particle selectors and kinematic region. An uncertainty of  $1.3\%$ is assigned to the background yield and $1.4\%$ to the signal efficiency. To determine the impact of the $\Lambda$ selection procedure, the $\Lambda$ yield is evaluated in the $\mes$ sideband region, using a 4-vector sum of $\proton$ and $\pi^-$ candidates to identify a $\Lambda$ control sample which is independent of the nominal kinematic fit procedure. The relative $\Lambda$ yields are determined from data and background MC, before and after applying the nominal $\Lambda$ selection to this sample, resulting in a $13\%$ correlated uncertainty on both the signal efficiency and background estimate.

The $E_{\rm extra}$ cut introduces a systematic uncertainty due to possible mis-modeling of low-energy clusters in simulation. To evaluate this, the cluster energies in the MC are scaled to match the $E_{\rm extra}$ distribution in data. Parametrically, the level of data--MC agreement in the $E_{\rm extra}$ distribution (see Fig.~\ref{fig4}) is found to be equivalent to applying a shift of $5$~MeV per cluster. This correction is applied to the MC and a systematic of $1.9\%$ for the signal efficiency and $11\%$ for the background estimate is assigned, corresponding to the full impact of this correction. Systematic uncertainties are summarized in Table~\ref{tab1}.

\begin{linenomath}
\begin{table}
\begin{center}
\caption{Summary of systematic uncertainties on the signal efficiency and backgrounds.}
\begin{tabular}{lcc}
\hline
\hline
Source & Signal efficiency & Background \\
\hline
\hline
Signal weighting &  9.6\%  & \\
MC modeling & 16\% & 17\%\\
Particle identification & 1.4\% & 1.3\%\\
$\Lambda$ selection & 13\%  & 13\%\\
$E_{\rm extra}$ & 1.9\% & 11\%\\
\hline
\hline
\end{tabular}
\label{tab1}
\end{center}
\end{table}
\end{linenomath}

\begin{figure}[ht]
\begin{center}
\includegraphics[width=\columnwidth]{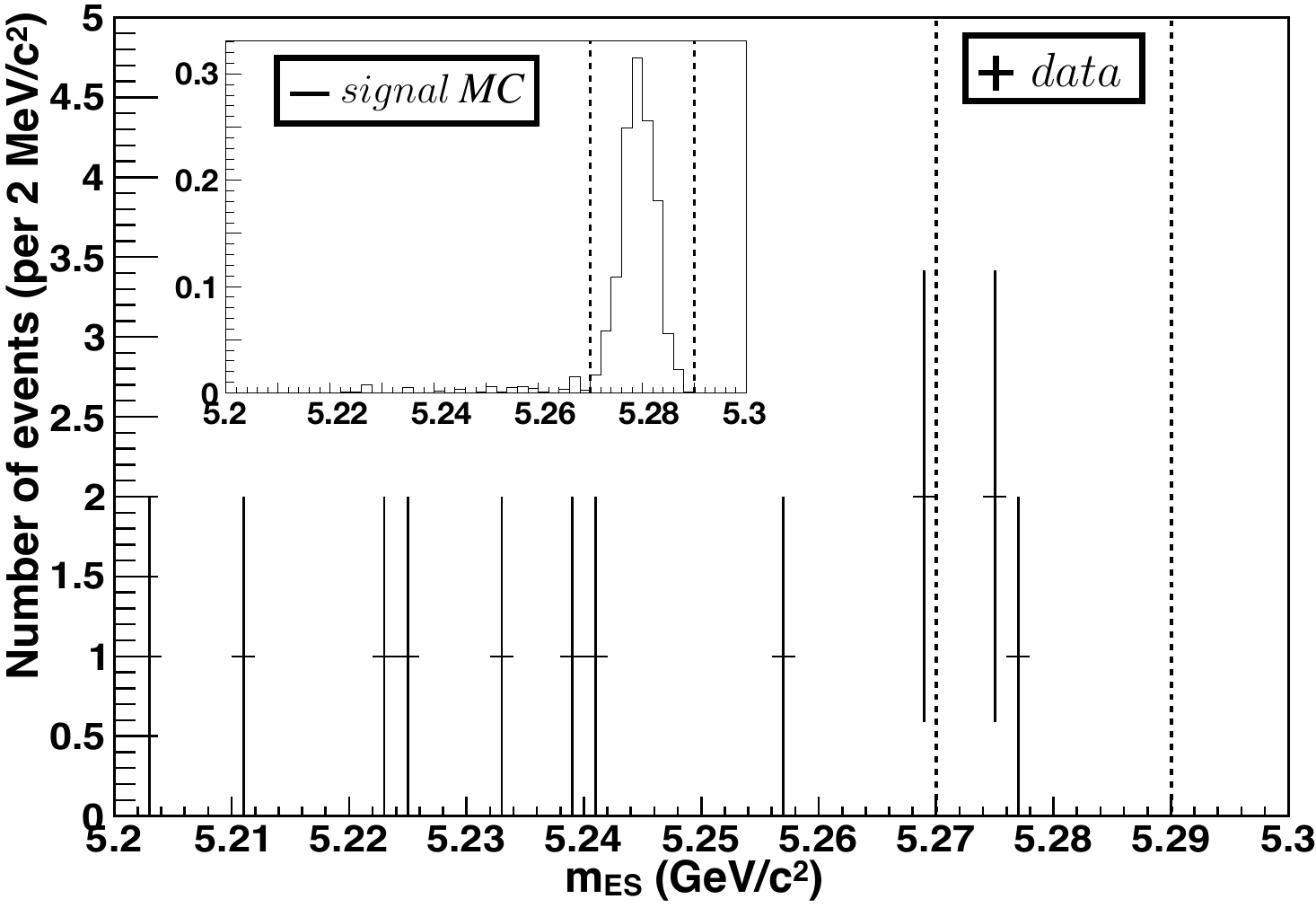}
\caption{The $\Btag$ $\mes$ distribution of events passing all other signal selection requirements for data and for signal MC (inset) scaled to a branching fraction of $\BFsignal$. The signal region is indicated by the vertical dashed lines, and the total background expected in the signal
region is $2.3 \pm 0.7~{\rm (stat.)} \pm 0.6~{\rm (sys.)}$ events.}
\label{fig6}
\end{center}
\end{figure}

\begin{figure}[ht]
\begin{center}
\includegraphics[width=\columnwidth]{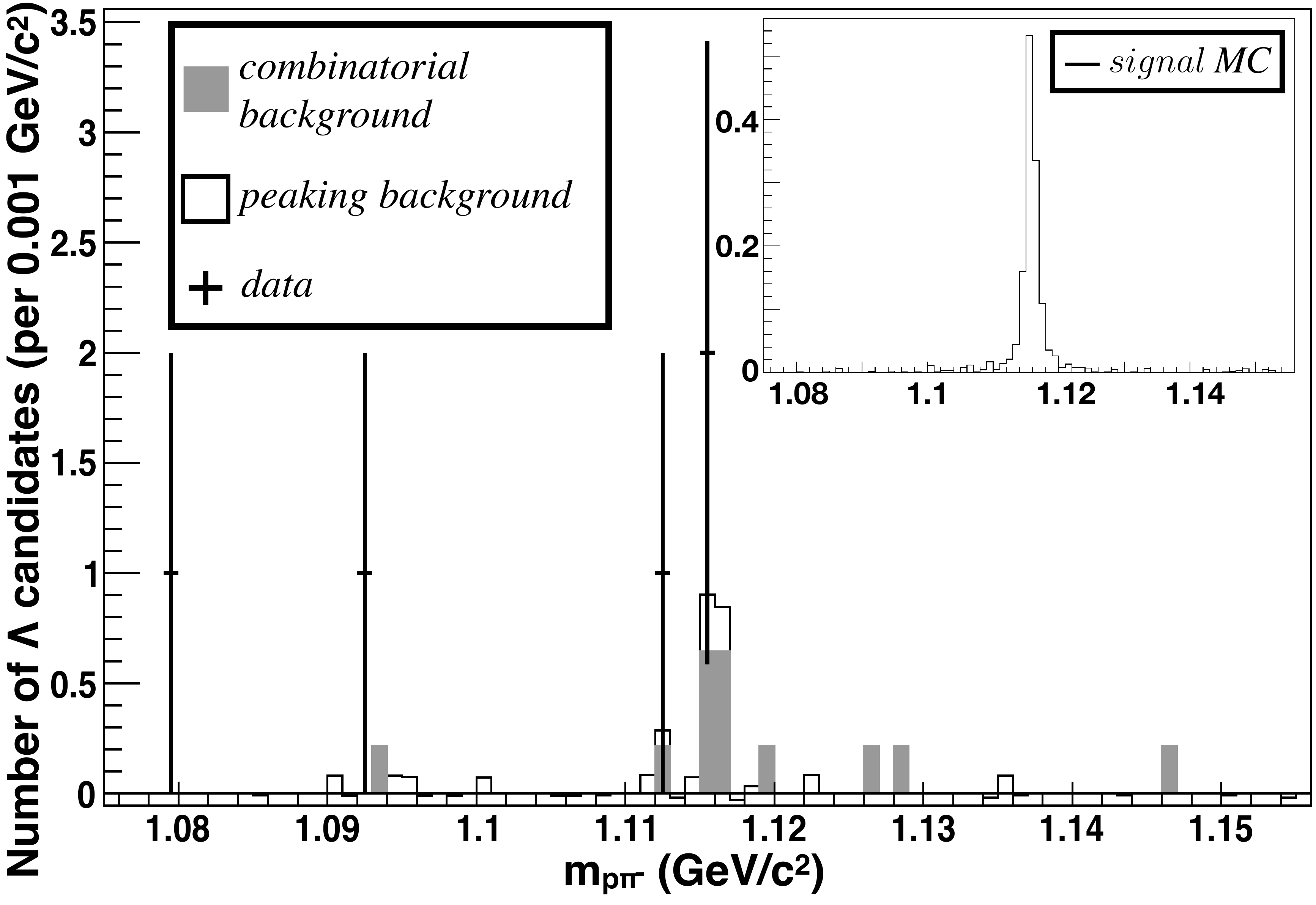}
\caption{The $\proton \pi^-$ invariant mass in events passing all other signal selection requirements. Data are shown as points with error bars, while the background expectation is shown as solid histograms. The negative bin values are a consequence of the background estimation procedure applied to low-statistics histograms. The expected signal distribution from MC is shown in the inset histogram, and is scaled to a branching fraction of  $\BFsignal$.}
\label{fig7}
\end{center}
\end{figure}

The $\Blpnn$ branching fraction is evaluated according to 
$ \BR (\Blpnn) = (N_{\rm data} - N_{\rm bg})/ (\epsilon^{\rm sig} \times N_{B^\pm}) \mbox{  ,}  $
 where $N_{\rm data}$ and $N_{\rm bg}$ are the number of events observed in data and the total estimated background yield, respectively. The overall $\Blpnn$ signal efficiency including the $\Lambda \to \proton \pi^-$ branching fraction \cite{pdg} is $\epsilon^{\rm sig} = (3.42 \pm 0.08 {\rm~(stat.)} \pm 0.80{\rm~(sys.)}) \times 10^{-4}$, and $N_{B^\pm}= \NBB$ is the estimated total number of charged $B$ mesons in the data sample~\cite{Bcount}. It is assumed that $\FourS \to \BB$ produces equal numbers of $\BzBzb$ and $\BpBm$ pairs. The selection efficiency is independent of $q^2$, the square of the four-momentum transfer to the $\nu\nub$ pair in signal events, within MC statistics.
A total of $N_{\rm data} = 3$ events are found in the $\mes$ signal region, consistent with the background yield expectation of $N_{\rm bg} = 2.3 \pm 0.7 {\rm~(stat.)} \pm 0.6 {\rm~(sys.)}$. The $\mes$ distribution of the $\Btag$ in events that pass all other selection requirements is plotted in Fig.~\ref{fig6}, and the $\proton \pi^-$ invariant mass distribution in Fig.~\ref{fig7}. The central value of the branching fraction is determined to be $\BFresult$. As no evidence is found for signal, a $90\%$ confidence level upper limit is computed using the Barlow method~\cite{barlow}, yielding $\BFlimit$.

A constraint can be placed on $|C^\nu_L|$, the Wilson coefficient that describes left-handed weak currents, by comparing this measurement to
the SM-predicted value.  Using the parametrization of Ref. \cite{Altmannshofer_2009}, and assuming the SM value of $C^\nu_R = 0$, 
a limit of $\epsilon \equiv |C^\nu_L|/|(C^\nu_L)^\textrm{SM}| <$  \epsilonLimLambda is obtained at the $90\%$ confidence level. 

In conclusion, a search has been performed for the FCNC decay process $\Blpnn$  based on the full $\babar$ dataset collected at the CM energy of the \FourS resonance. No evidence is found for an excess over the SM prediction and the first branching fraction limit on this decay is reported.

\section{Acknowledgments}
\label{sec:Acknowledgments}
The authors gratefully acknowledge David M. Straub's help in determining the new physics implications of this analysis.
We are grateful for the excellent luminosity and machine conditions
provided by our \pep2\ colleagues, 
and for the substantial dedicated effort from
the computing organizations that support \babar.
The collaborating institutions wish to thank 
SLAC for its support and kind hospitality. 
This work is supported by
DOE
and NSF (USA),
NSERC (Canada),
CEA and
CNRS-IN2P3
(France),
BMBF and DFG
(Germany),
INFN (Italy),
FOM (The Netherlands),
NFR (Norway),
MES (Russia),
MICIIN (Spain),
STFC (United Kingdom). 
Individuals have received support from the
Marie Curie EIF (European Union),
the A.~P.~Sloan Foundation (USA),
and the Binational Science Foundation (USA-Israel).

\end{document}